\documentclass[notitlepage,superscriptaddress,longbibliography,nofootinbib,twocolumn,floatfix]{revtex4-2}

\usepackage[utf8]{inputenc}
\usepackage{braket}
\usepackage{amsmath,amsthm,amssymb,amsfonts}
\usepackage{mathtools}
\usepackage{graphicx}
\usepackage{hyperref}
\usepackage{comment}
\usepackage{color}
\usepackage{makecell}
\usepackage{enumitem}
\usepackage{lineno}

\usepackage[dvipsnames]{xcolor}
\usepackage{hyperref}
\usepackage{braket}
\usepackage{bbm}
\usepackage{bm}
\usepackage{longtable}
\usepackage[bb=boondox]{mathalfa}
\usepackage{adjustbox}
\usepackage{array}
\usepackage{multirow}
\usepackage{tabularx}
\usepackage{float}
\usepackage{svg}
\usepackage{adjustbox}

\newtheorem{proposition}{Proposition}

\usepackage{array}

\usepackage[normalem]{ulem} 

\definecolor{dred}{rgb}{.9,.1,0}
\definecolor{dorange}{rgb}{.85,.3,0}
\definecolor{dgreen}{rgb}{0,.8,0}

\usepackage{tikz}
\usepackage{tikz,esvect,tikz-3dplot,calc}
\usetikzlibrary{3d,calc,intersections,arrows.meta}
\usetikzlibrary{decorations.pathreplacing}
\usetikzlibrary{decorations.markings}
\tikzset{->-/.style={decoration={
			markings,
			mark=at position #1 with {\arrow{>}}},postaction={decorate}}}

\newcommand{\ccaption}[2]{\caption[#1]{\textit{#1.} #2}}

\newcommand{\ii}{\mathrm{i}}
\newcommand{\ie}{{\it i.e.},\ }
\newcommand{\eg}{{\it e.g.},\ }
\newcommand{\id}{\mathbb{1}} 
\newcommand{\norm}[1]{\left\lVert#1\right\rVert}

\begin{document}
	\title{Experimental proposal to probe the extended Pauli principle}
	
	\author{Lucas Hackl}
	\thanks{Equal contribution}
	\email{lucas.hackl@unimelb.edu.au}
	\affiliation{School of Mathematics and Statistics \& School of Physics, The University of Melbourne, VIC 3010, Parkville, Australia}
	\affiliation{Department of Mathematical Sciences, University of Copenhagen, Universitetsparken 5, 2100, Copenhagen, Denmark}
	
	\author{Dayang Li}
	\thanks{Equal contribution}
	\email{dayli@dtu.dk}
	\affiliation{Department of Electrical and Photonics Engineering, Technical University of Denmark, Ørsteds Plads Building 343, 2800, Kongens Lyngby, Denmark}
	
	\author{Nika Akopian}
	\email{nikaak@dtu.dk}
	\affiliation{Department of Electrical and Photonics Engineering, Technical University of Denmark, Ørsteds Plads Building 343, 2800, Kongens Lyngby, Denmark}
	
	\author{Matthias Christandl}
	\email{christandl@math.ku.dk}
	\affiliation{Department of Mathematical Sciences, University of Copenhagen, Universitetsparken 5, 2100, Copenhagen, Denmark}
	
	\begin{abstract}
		All matter is made up of fermions --- one of the fundamental type of particles in nature. Fermions follow the Pauli exclusion principle, stating that two or more identical fermions cannot occupy the same quantum state. Antisymmetry of the fermionic wavefunction, however, implies additional constraints on the natural occupation numbers. These constraints depend on the dimensionality and purity of the system and have so far not been explored experimentally in a fermionic system. Here, we propose an experiment in a multi-quantum-dot system capable of producing the highly entangled fermionic states necessary to reach the regime, where these additional constraints become dominant and can be probed. The type and strength of the required multi-fermion entanglement provides barriers to reaching deep into this regime. Transcending these barriers thus serves as a testing ground for the capabilities of future fermionic quantum information processing as well as quantum computer architectures based on fermionic states. All operations in our proposal are based on all-optical gates presented in [arXiv:2107.05960]. We simulate our state preparation procedures in realistic structures, including all main decoherence sources, and find fidelities above 0.97.
	\end{abstract}

	\maketitle
	
	\section{Introduction}
	Pauli's exclusion principle~\cite{pauli1925einfluss} is a fundamental property of quantum mechanics and can be stated plainly without direct reference to quantum theory by saying that two identical fermionic particles cannot occupy the same state, \eg location. Its implications are far reaching and include both the elaborate electron shell structure of atoms leading to the variety of chemical elements and the stability of matter~\cite{dyson1967stability} itself including neutron stars.
	
\begin{figure*}[t]
		\centering
		\includegraphics[width=\linewidth]{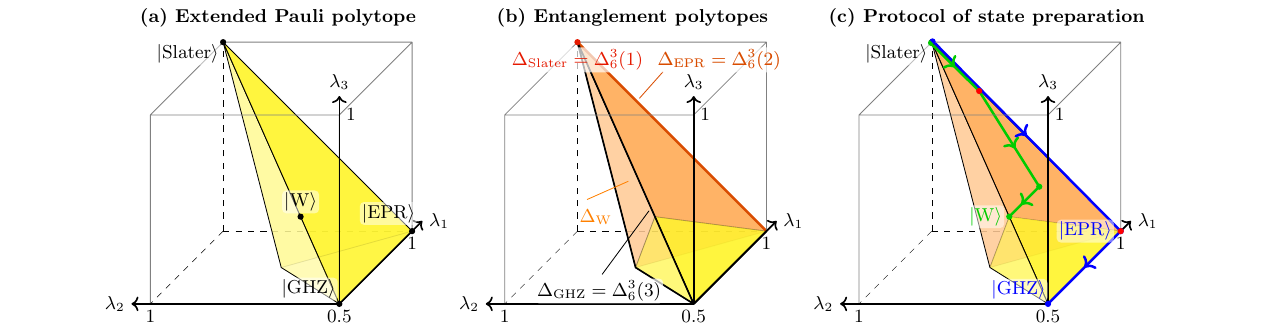}
		\ccaption{Natural occupation numbers for Borland-Dennis setup ($N=3$, $d=6$)}{\textbf{(a)} We show the \emph{extended Pauli polytope} $\Delta^3_6$ consisting of the natural occupation numbers $(\lambda_1,\lambda_2,\lambda_3)\in\mathbb{R}^3$ with $\lambda_1\geq\lambda_2\geq\lambda_3$ satisfying the Borland-Dennis inequality~\eqref{eq:schematic}. Note that have $\vec{\lambda}=(\lambda_1,\dots,\lambda_6)$, but it was shown in~\cite{borland1972conditions} that $\lambda_1+\lambda_6=\lambda_2+\lambda_5=\lambda_3+\lambda_4=1$, so we can pick $(\lambda_1,\lambda_2,\lambda_3)$ as the only independent variables. We also indicate the positions of the characteristically entangled states from Table~\ref{tab:states}. \textbf{(b)} The system has four distinct entanglement classes $\mathcal{C}$, which we label by their four characteristically entangled states from Table~\ref{tab:states}. We show the associated entanglement polytopes: $\Delta_{\mathrm{Slater}}$ (red top vertex), $\Delta_{\mathrm{EPR}}$ (orange edge on the $\lambda_1=1$ plane), $\Delta_{\mathrm{W}}$ (orange/darker polytope) and $\Delta_{\mathrm{GHZ}}$ (full polytope). Three of these polytopes coincide with the respective polytopes $\Delta^3_6(m)$ of (at most) $m$-fermion entangled states (for $m=1,2,3$), \ie we also probe how many fermions are genuinely entangled. \textbf{(c)} Our experimental proposal is based on the preparation of highly entangled states by applying certain gates to an unentangled Slater determinant state. We indicate the two main protocols that allow us to prepare the states $\ket{\mathrm{W}}$ (endpoint of the green/lighter path) or $\ket{\mathrm{EPR}}$ and $\ket{\mathrm{GHZ}}$ (turning point and endpoint of the blue/darker path) out of the unentangled state $\ket{\mathrm{Slater}}$. We indicate where we transcend the barriers of a given entanglement polytope (red dots at the first turning point of the green/lighter path and the blue/darker path respectively).
		}
		\label{fig:Borland-Dennis}
	\end{figure*}

	The \emph{extended Pauli principle} further restricts the possible pure fermionic quantum states. An important implication is the pinned state effect~\cite{klyachko2009pauli,klyachko2013pauli,schilling-pinning}, when the dynamics of a quantum system drives the system to the boundary of allowed states and turns the extended Pauli constraints into equalities. This observation may also lead to an interesting generalization of the Hartree-Fock method~\cite{schilling-pinning}. Viewing the Pauli contraint as a constraint on the $1$-body reduced density matrix (1-RDM, see~\eqref{eq:1-body}), the 1960s saw activity in the study of further constraints on reduced density matrices (see \eg \cite{coleman1963structure,ruskai1969n,coleman2000reduced}). In 1972, Borland and Dennis~\cite{borland1972conditions,ruskai2007connecting} discovered that for three fermions in six modes in a pure state, already the 1-RDM has further constraints, best captured as restrictions on its ordered eigenvalues $\lambda_i$. These are the \emph{natural occupation numbers}, \ie the occupation numbers measured in the basis, where the 1-RDM is diagonal. Surprisingly, the constraints could be determined exactly and are given by a set of linear inequalities, \eg
	\begin{equation}
		\lambda_1 + \lambda_2+\lambda_4\leq 2\,,\label{eq:schematic}
	\end{equation}
	inscribing a polytope (Figure~\ref{fig:Borland-Dennis}). The six natural occupation numbers are in decreasing order and therefore~\eqref{eq:schematic} goes beyond the Pauli principle $0\leq\lambda_i \leq 1$, where we recall $\sum^6_{i=1} \lambda_i=N=3$. In 2008, Altunbulak and Klyachko~\cite{altunbulak2008pauli} understood that the polytopal nature was no accident but a consequence of the symplectic geometric structure of the problem and were therefore able to give a general algorithm to determine the inequalities of the extended Pauli principle for general $N$ and $d\leq 10$. The situation parallels that for spin systems, where it is known as the quantum marginal problem~\cite{christandl2006spectra,klyachko2004quantum,DAFTUAR200580,christandl2007nonzero}. The extended Pauli principle would show its relevance when its inequalities are saturated or near-saturation. This is where the extended Pauli principle adds real constraints to the system, that are overlooked (\ie not imposed) by the regular Pauli principle. This situation has recently been investigated in a variety of systems both analytically and numerically~\cite{klyachko2009pauli,klyachko2013pauli,schilling-pinning,reiher,schilling2015hubbard,schilling2015quasipinning,benavides2019time,schilling2020implications}. It was also simulated on the IBM Quantum Experience, where the fermionic system was mapped to qubits~\cite{smart2019experimental}. Here, we propose for the first time to probe the extended Pauli principle in a genuine fermionic system and to benchmark the ability of a given fermionic quantum technology to produce strongly correlated fermionic states.
	
	\begin{table*}[t!]
		\centering
		\renewcommand{\arraystretch}{1.3}
		\begin{tabular}{lllll}
			\textbf{State} & \textbf{First quantization} & \textbf{Second quantization} & $\vec{\lambda}$ & \textbf{$E(\ket{\Psi_N})$} \\
			\hline
			$\ket{\mathrm{Slater}}$ & $\ket{e_1}\wedge\ket{e_2}\wedge\ket{e_3}$ & $\ket{101010}$ & $(1,1,1,0,0,0)$ & $\log{3}\approx1.099$ \\
			$\ket{\mathrm{EPR}}$ & $\tfrac{1}{\sqrt{2}}(\ket{e_1}\wedge\ket{e_2}+\ket{e_4}\wedge\ket{e_5})\wedge\ket{e_3}$ & $\tfrac{1}{\sqrt{2}}(\ket{101010}+\ket{010110})$ & $(1,\tfrac{1}{2},\tfrac{1}{2},\tfrac{1}{2},\tfrac{1}{2},0)$ & $\frac{1}{3}\log{108}\approx 1.561$ \\
			$\ket{\mathrm{W}}$ & $\tfrac{1}{\sqrt{3}}(\ket{e_1}\!\!\wedge\!\!\ket{e_2}\!\!\wedge\!\!\ket{e_3}\!\!+\!\!\ket{e_4}\!\!\wedge\!\!\ket{e_5}\!\!\wedge\!\!\ket{e_3}\!\!+\!\!\ket{e_2}\!\!\wedge\!\!\ket{e_4}\!\!\wedge\!\!\ket{e_6})$ & $\tfrac{1}{\sqrt{3}}(\ket{101010}+\ket{010110}+\ket{011001})$ & $(\tfrac{2}{3},\tfrac{2}{3},\tfrac{2}{3},\tfrac{1}{3},\tfrac{1}{3},\tfrac{1}{3})$ & $\frac{2}{3}\log\frac{27}{2}\approx1.735$ \\
			$\ket{\mathrm{GHZ}}$ & $\tfrac{1}{\sqrt{2}}(\ket{e_1}\wedge\ket{e_2}\wedge\ket{e_3}+\ket{e_4}\wedge\ket{e_5}\wedge\ket{e_6})$ & $\tfrac{1}{\sqrt{2}}(\ket{101010}+\ket{010101})$ & $(\tfrac{1}{2},\tfrac{1}{2},\tfrac{1}{2},\tfrac{1}{2},\tfrac{1}{2},\tfrac{1}{2})$ & $\log{6}\approx1.792$
		\end{tabular}
		\ccaption{Characteristically entangled states}{We represent the four states in first and second quantization, list their natural occupation numbers $\vec{\lambda}$ and calculate their fermionic quantum functional $E(\ket{\Psi_N})$ from~\eqref{eq:lambda-state}. The states $\ket{e_i}$ form a basis of the one-fermion Hilbert space $\mathcal{H}_6$, such that $\ket{e_i}=\hat{a}_i^\dagger\ket{0}$. \ie we have $\ket{n_1,\dots,n_6}=(\hat{a}_1^\dagger)^{n_1}\dots (\hat{a}_6^\dagger)^{n_6}\ket{0}$. The natural occupation numbers $\vec{\lambda}$ are the eigenvalues of the 1-RDM.}
		\label{tab:states}
	\end{table*}
	
	\section{Three probes of the Extended Pauli Principle}\label{sec:three-probes}
	In order to probe the extended Pauli principle, we need to map out the parameter space of $\lambda_i$ inside the constraints and in particular identify the regime closest to saturation. For this, we propose \emph{three probes} that give rise to an onion-like structure of successively stronger inequalities, that we can study experimentally and of which the extended Pauli principle forms the outermost layer. These three probes consist of \emph{(A) entanglement polytopes}, \emph{(B) $m$-fermion entanglement} and \emph{(C) fermionic quantum functional}. We can then experimentally demonstrate that we reached the outermost layer by explicitly violating all the stronger inequalities, such that only the weakest and thus most general inequalities (extended Pauli principle) remain satisfied, as confirmed by all three of our probes.
	
	We consider a system of $N$ fermions in a system with $d\geq N$ modes. The one-particle Hilbert space is then $\mathcal{H}_d$ with $\dim \mathcal{H}_d=d$, while the resulting $N$-particle Hilbert space is $\Lambda^N\mathcal{H}_d$, spanned by completely anti-symmetrized tensor products $\ket{e_{i_1}}\wedge\dots\wedge \ket{e_{i_N}}$ of $N$ elements $\ket{e_{i}}\in\mathcal{H}_d$.
	
	We can also describe pure quantum states $\ket{\Psi_N}$ in second quantization using $d$ fermionic creation and annihilation operators $\hat{a}_i^\dagger$ and $\hat{a}_i$, which satisfy the canonical anti-commutation relations $\{\hat{a}_i,\hat{a}_j^\dagger\}=\delta_{ij}$ with vacuum state $\ket{0}$, such that $\hat{a}_i\ket{0}=0\,\forall\,i$. This allows us to write $\ket{e_i}=\hat{a}_i^\dagger\ket{0}$ and more generally $\ket{e_{i_1}}\wedge\dots\wedge\ket{e_{i_N}}=\hat{a}_{i_1}^\dagger\dots\hat{a}^\dagger_{i_N}\ket{0}$. The 1-RDM
	\begin{linenomath*}\begin{align}
			\gamma_{ij}=\frac{\braket{\Psi_N|\hat{a}_j^\dagger \hat{a}_i|\Psi_N}}{\braket{\Psi_N|\Psi_N}}\,,\label{eq:1-body}
	\end{align}\end{linenomath*}
	of a pure state $\ket{\Psi_N}\in\Lambda^n\mathcal{H}_d\subset \mathcal{H}$ is a Hermitian $d$-by-$d$ matrix and satisfies $\mathrm{Tr}(\gamma)=N$. It can be diagonalized by a unitary transformation $U$, such that
	\begin{linenomath*}\begin{align}
			U\gamma U^\dagger=\mathrm{diag}(\lambda_1,\dots,\lambda_d)\,,
	\end{align}\end{linenomath*}
	where we choose the order $\lambda_1\geq\lambda_2\geq\dots\geq\lambda_d$. We refer to $\vec{\lambda}=(\lambda_1,\dots,\lambda_d)$ as the \emph{natural occupation numbers} of $\ket{\Psi_N}$. They are the expectation values of the transformed number operators $\hat{n}_i'=\hat{a}_i'^{\dagger}\hat{a}'_i$ with $\hat{a}'_i=\sum_j U_{ij}\hat{a}_j$.
	
	\subsection{Entanglement polytopes}\label{sec:e-poly}
	The entanglement class $\mathcal{C}\subset\mathcal{H}$ consists of all pure states that can be converted into each other with finite probability of success using local operations and classical communication (stochastic LOCC = SLOCC~\cite{mathonet2010entanglement}). For a fermionic system with $d$ modes, this is implemented by a representation of the group $\mathrm{GL}(d)$ (see~\cite{walter2013entanglement}). For every class $\mathcal{C}$, the entanglement polytope $\Delta_{\mathcal{C}}\subset\mathbb{R}^d$ is defined as the natural occupation numbers\footnote{If we had $N$ distinguishable particles, we would have $N$ vectors $\vec{\lambda}^{i}$ consisting of the eigenvalues of the $N$ different 1-RDMs $\rho^{(i)}$, as explained in~\cite{walter2013entanglement}.} $\vec{\lambda}$  of states $\ket{\Psi_N}$ in the closure $\overline{\mathcal{C}}$, which is shown to be a convex polytope  in~\cite{walter2013entanglement}. The notion of entanglement polytopes is coarser than the one of entanglement classes, \ie the entanglement polytopes of two different entanglement classes $\mathcal{C}\neq \mathcal{C}'$ may be identical, although it is possible that when considering the SLOCC conversion of many copies of a state, the polytope contains all relevant information. We have the clear-cut criterion that
	\begin{linenomath*}\begin{align}
			\vec{\lambda}\neq\Delta_{\mathcal{C}}\quad\Rightarrow\quad\ket{\Psi_N}\neq\mathcal{C}\,,
	\end{align}\end{linenomath*}
	\ie if $\vec{\lambda}$ is not in the polytope $\Delta_{\mathcal{C}}$, the state must be in another entanglement class. Entanglement polytopes form a hierarchy, where $\Delta_{\mathcal{C}'}\subset\Delta_{\mathcal{C}}$ if a state $\ket{\psi'}\subset\mathcal{C}'$ can be approximated by a sequence of states contained in $\mathcal{C}$.
	
	\begin{figure*}[t!]
		\centering
		\includegraphics[width=\linewidth]{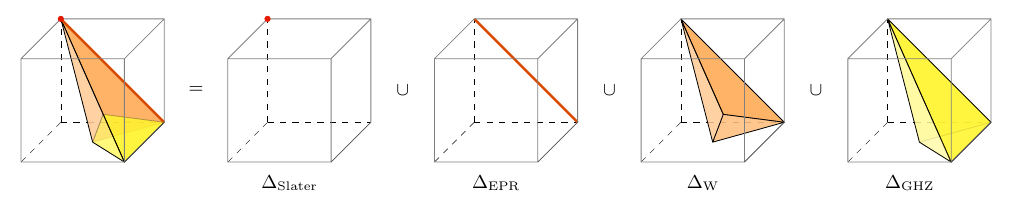}
		\ccaption{Borland-Dennis entanglement polytopes}{For $N=3, d=6$, we have four distinct entanglement classes $\mathcal{C}$ with associated entanglement polytopes $\Delta_\mathcal{C}$. They are in one-to-one correspondence to the four characteristic states.}
		\label{fig:DB-entanglement-polytopes}
	\end{figure*}
	
	The polytope constraints for Borland-Dennis in figure~\ref{fig:DB-entanglement-polytopes} are
	\begin{linenomath*}\begin{align}
			\begin{split}
				\mathrm{Slater:}&\quad \lambda_1=\lambda_2=\lambda_3=1\,,\\
				\mathrm{EPR:}&\quad \lambda_1=1\,,\,\lambda_2=\lambda_3\,,\\
				\mathrm{W:}&\quad \lambda_1+\lambda_2-\lambda_3\leq 2\,,\,\lambda_1+\lambda_2+\lambda_3\geq 2\,,\\
				\mathrm{GHZ:}&\quad \lambda_1+\lambda_2-\lambda_3\leq 2\,,\\
			\end{split}  \label{eq:polytope-ineq}
	\end{align}\end{linenomath*}
	where we have $1\geq \lambda_1\geq\dots\geq\lambda_3\geq\tfrac{1}{2}$, $\lambda_{4}=1-\lambda_3$, $\lambda_{5}=1-\lambda_2$ and $\lambda_{6}=1-\lambda_1$.
	
	\subsection{Polytopes of (at most) \texorpdfstring{$m$}{m}-fermion entangled states}\label{sec:m-fermion}
	A state $\ket{\Psi_N}$ is \emph{(at most) $m$-fermion entangled} if
	\begin{linenomath*}\begin{align}
			\ket{\Psi_N}=\ket{\psi_m}\wedge \ket{\psi_1^{(1)}}\wedge\dots\wedge\ket{\psi^{(N-m)}_{1}}\,,\label{eq:n-particle-entangled}
	\end{align}\end{linenomath*}
	\ie such a state consists of a single $m$-fermion state $\ket{\psi_m}\in\Lambda^m(\mathcal{H}_d)$ wedged with $(N-m)$ single-fermion states in $\mathcal{H}_d$. We refer to this set as $\mathcal{C}^N_d(m)\subset\Lambda^N(\mathcal{H}_d)$. A state $\ket{\Psi_N}$ is \emph{genuinely $m$-fermion entangled} if $\ket{\Psi_N}\in \mathcal{C}^N_d(m)\setminus \mathcal{C}^N_d(m-1)$, \ie it is (at most) $m$-fermion entangled, but not $(m-1)$-fermion entangled. We define the polytope $\Delta^N_d(m)$ as the set of $\vec{\lambda}\in\mathbb{R}^d$ associated to states in the closure of $\mathcal{C}^N_d(m)$. In particular, we have $\Delta^N_d(N)=\Delta^N_d$ forming the full extended Pauli polytope. This leads to the following simple proposition~\cite{klyachko2009pauli,reuvers2021generalized}:
	
	\begin{figure}[t]
		\centering
		\includegraphics[width=\linewidth]{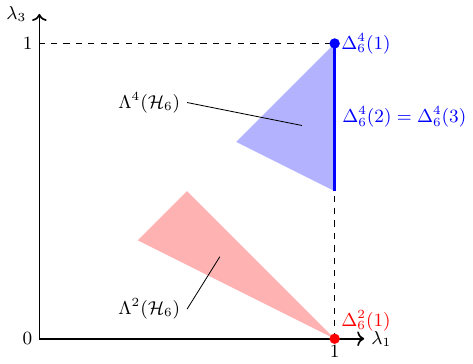}
		\ccaption{Genuine $m$-fermion entanglement}{We show for $\Lambda^2(\mathcal{H}_6)$ in red and $\Lambda^4(\mathcal{H}_6)$ in blue the (at most) $m$-fermion entangled polytopes $\Delta^N_6(m)$. Using $\vec{\lambda}=(\lambda_1,\lambda_1,\lambda_3,\lambda_3,2-\lambda_1-\lambda_3,2-\lambda_1-\lambda_3)$, we can illustrate these polytopes in a single figure in 2d. In particular, we find $\Delta^4_6(2)=\Delta^4_6(3)$. Note that the respective (at most) $m$-fermion entangled polytopes for $\Delta^3_6(m)$ were already shown in figure~\ref{fig:DB-entanglement-polytopes}, where we saw that $\Delta^3_6(1)=\Delta_{\mathrm{Slater}}$, $\Delta^3_6(2)=\Delta_{\mathrm{EPR}}$ and $\Delta^3_6(3)=\Delta_{\mathrm{GHZ}}$.
		}
		\label{fig:n-particle-entanglement}
	\end{figure}
	
	\begin{proposition}
		The natural occupation numbers $\vec{\lambda}$ of a state $\ket{\Psi_N}\in\mathcal{C}^N_d(m)$ are
		\begin{linenomath*}\begin{align}
				(\hspace{-2mm}\underbrace{1,\dots,1}_{(N-m)\text{ times}}\hspace{-2mm},\lambda_{N-m+1},\dots,\lambda_{d})\,,
		\end{align}\end{linenomath*}
		where the last $m$ natural occupation numbers satisfy the conditions of the extended Pauli principle for a general state $\ket{\Psi_{d-(N-m)}}\in \Lambda^{m}(\mathcal{H}_{d-(N-m)})$.
	\end{proposition}
	\begin{proof}
		We consider a state of the form~\eqref{eq:n-particle-entangled} and write the $1$-particle states as $\ket{\psi_1^{(k)}}=\sum_i v^{(k)}_i\hat{a}_i^\dagger\ket{0}$. We assume that all the states $\ket{\psi_1^{(k)}}$ and $\ket{\psi_m}$ are independent, \ie their wedge product does not vanish (otherwise $\ket{\Psi}=0$). We then choose mutually orthogonal $\ket{\phi_1^{(k)}}$, such that $\ket{\psi_1^{(k)}}\wedge\dots\wedge\ket{\psi_1^{(k)}}=\ket{\phi_1^{(k)}}\wedge\dots\wedge\ket{\phi_1^{(k)}}$. This allows us to define $N$ new creation operators $\hat{b}^\dagger_k=\sum_j U_{kj}\hat{a}_k$ (with unitary matrix $U$), such that for the first $N-m$ ones we have $\ket{\phi_1^{(k)}}\propto\hat{b}^\dagger_i\ket{0}$. We then have $\ket{\Psi_N}\propto\hat{b}_1^\dagger\dots\hat{b}^\dagger_{N-m}\ket{\psi_n}$. The 1-RDM with respect to $\hat{b}_k$ is block diagonal with $(N-m)$ ones on the diagonal, because $\ket{\Psi_N}$ is an eigenstate of the first $(N-m)$ number operators $\hat{n}_i=\hat{b}_i^\dagger\hat{b}_i$. The remaining $d-(N-m)$ quadratic block describes then $n$ particles in a system with $d-(N-m)$ fermionic modes, such that the respective eigenvalues $\lambda_{N-m+1}$ up to $\lambda_{d}$ satisfy the conditions of $\Lambda^{m}(\mathcal{H}_{d-(N-m)})$.
	\end{proof}
	
	We illustrate $\Lambda^2(\mathcal{H}_6)$ and $\Lambda^4(\mathcal{H}_6)$ in figure~\ref{fig:n-particle-entanglement}.

	\subsection{Fermionic quantum functional}
	While the Slater determinant is the least entangled fermionic state (satisfying the most restrictive inequalities), probing the extended Pauli principle requires the preparation of the most entangled fermionic states. Since we focus on pure fermionic states, quantifying their local mixedness is a good way to measure fermionic entanglement just as it is for spin states. The analogue of the Schmidt rank for bipartite spin state (a SLOCC monotone) becomes the number of non-zero natural occupation numbers (a fermionic SLOCC monotone). In~\cite{Christandl_2021}, the quantum functionals, a large family of generalization of Schmidt rank to multiparticle entangled spin states have been introduced as optimizations over an entanglement polytope. The (logartithm of a) quantum functionals are the first additive entanglement measures. A bosonic variant has been studied in~\cite{christandl2021communication}, while we propose the logarithm of the \emph{fermionic quantum functional}
	\begin{linenomath*}\begin{align}
			\hspace{-2mm}    E(\ket{\Psi_N})=\max_{\vec{\lambda}\in\Delta_{\Psi_N}} H(\bar{\lambda})\,\,\,\text{with}\,\,\,
			H(\bar\lambda)=-\sum^d_{i=1} \bar\lambda_i\log\bar\lambda_i\label{eq:lambda-state}
	\end{align}\end{linenomath*}
	being the Shannon entropy of the probability distribution $\bar\lambda=\frac{\lambda}{N}:=(\frac{\lambda_1}{N}, \ldots, \frac{\lambda_d}{N})$. The maximization over the entanglement polytope $\Delta_{\Psi_N}$ of $\ket{\Psi_N}$ implies that $E(\ket{\Psi_N})$ is constant for all pure states in an entanglement class and even more generally for all states with the same entanglement polytope. The fermionic quantum functional measures the distance\footnote{This can be made rigorous when reformulating it as a relative entropy distance.} to the most locally mixed state, which, if it exists (which it may not for large $N$, see remark 12 in~\cite{reuvers2021generalized}), has natural occupations $\lambda_i=\frac{N}{d}$ and corresponding functional value $E=\log d$. A Slater determinant in contrast has the smallest value $E=\log N$, since $N \leq d$.
	
	\begin{figure}[!ht]
		\includegraphics[width=\linewidth]{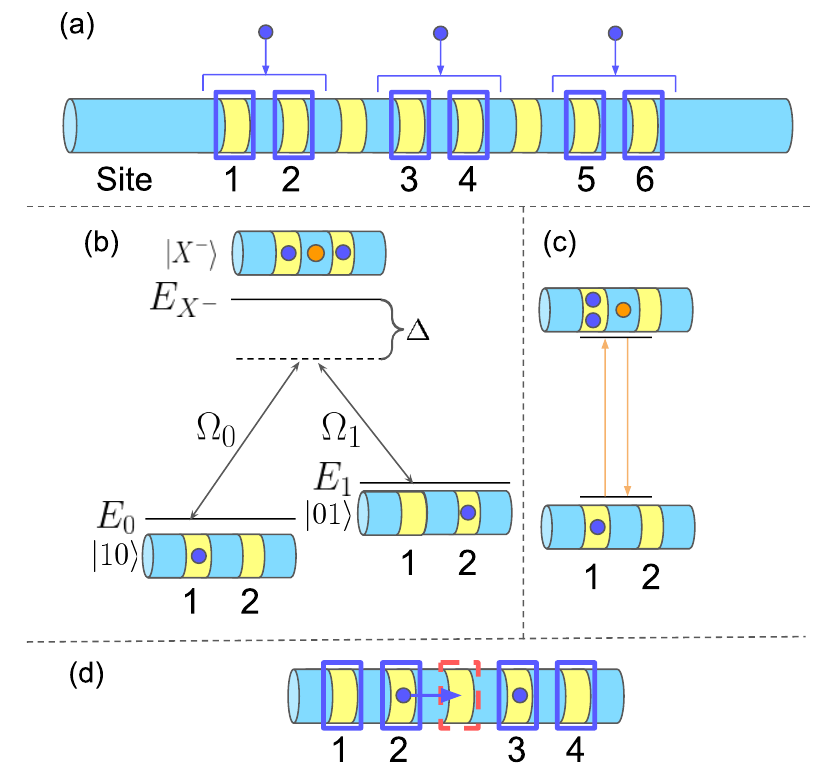}
		\ccaption{The physical system to test the Borland-Dennis relations, namely the crystal-phase quantum dots in a semiconductor nanowire}{\textbf{(a)} A nanowire with eight crystal-phase quantum dots for electrons (yellow sections). We will focus on six of the sites. Three electrons (blue dots) are distributed among this six sites, one electron per two sites. \textbf{(b)} The coherent manipulation scheme~\cite{li2021locationqubit}. The $\Lambda$-system connecting neighbouring quantum dots. Two location states $\ket{10}$ and $\ket{01}$ share a common excited state $\ket{X^-}$ with two electrons, one on each site, and a hole (orange dot) that allows for the coupling of the two states. An optical scheme utilizing coherent population trapping in a three level system can be used to coherently manipulate the system~\cite{li2021locationqubit}. \textbf{(c)} The site occupation measurement scheme. Since the crystal-phase quantum dots can be fabricated according to design~\cite{harmand2018step}, the confinement potential for the electrons and holes can be engineered. Consequently their energy levels can also be tuned such that only the transition shown can be resonantly driven (orange arrows). Other location states are unaffected. Spectral signatures from time resolved photoluminescence measurements will differentiate the $\ket{10}$ and $\ket{01}$ states, hence probe the occupation of a site. \textbf{(d)} The mechanism behind the conditional location state manipulations. If a location state $\ket{01}$ is present on sites 1-2, the electron on site 2 can be transferred to the auxiliary site (red dashed box). Due to the presence of an electron on the auxiliary site, the transition energies on sites 3-4 will be modified due to Coulomb interaction~\cite{li2021locationqubit}. Hence by applying pulses according to the modified energy levels, one can perform conditional location state manipulations on sites 3-4.}
		\label{fig:STIRSAPLocQb}
	\end{figure}
	
	\section{Experimental proposal}
	\label{sec:expProp}
	The physical platform to probe the Borland-Dennis relations is required to be a genuinely fermionic system. Crystal-phase quantum dots~\cite{akopian2010crystal} on a III-V semiconductor nanowire (for instance GaAs or InP) is one such system. The crystal structure switches between wurtzite and zincblende (aka sphalerite) along the nanowire~\cite{akopian2010crystal,harmand2018step, lehmann2013sharp}, resulting in type-II band alignment, allowing electrons (holes) to be confined in the zincblende (wurtzite) sections, hence the term crystal-phase quantum dots. If we consider a wurtzite section sandwiched between two zincblende sections, two electronic states (our fermionic states), one on each zincblende section, can be coupled by a shared hole in the wurtzite section~\cite{li2021locationqubit,hastrup2020charging} (Figure~\ref{fig:STIRSAPLocQb}(b)). We fully spin-polarize the electrons by fixing the driving field polarizations at one particular circular polarization throughout initialization and manipulation, so that we operate with electrons with only one specific spin.

	Placing three electrons in a system of six sites (quantum dots) provides an exact realization of the Borland-Dennis setup (Figure~\ref{fig:STIRSAPLocQb}(a)). It suffices for our probes to study states with one electron per sites 1-2, 3-4, and 5-6, but in principle we can prepare any state of the system, for instance states with electrons occupying sites 1-2. For coherent manipulation we move an electron between sites, create a superposition of its location, and entangle neighbouring electrons' locations~\cite{li2021locationqubit}. Focusing on sites 1-2, an electron occupies site 1 (2) in the location state $\ket{10}$ ($\ket{01}$) (Figure~\ref{fig:STIRSAPLocQb}(b)). The states $\ket{10}$ and $\ket{01}$ are both coupled to a common excited state $\ket{X^-}$. This is only possible because they share a hole state. The system is driven by two optical fields represented by Rabi frequencies $\Omega_0$ and $\Omega_1$, enabling coherent population trapping and transfer~\cite{caillet2007precision,li2021locationqubit} using an adiabatic optical scheme. When the system is driven, three eigenstates are formed, two of which are near degenerate ($\ket{\Phi_1(0)}$ and $\ket{\Phi_2(0)}$). They almost do not contain any negatively charged exciton state component, and are effectively linear combinations of $\ket{10}$ and $\ket{01}$. Therefore, an arbitrary initial state $\ket{\psi}$ can be decomposed into a combination of the two eigenstates $\ket{\Phi_1(0)}$ and $\ket{\Phi_2(0)}$ with a small, yet finite energy difference. As a result, when the driving fields are active, the $\ket{\Phi_1(0)}$ and $\ket{\Phi_2(0)}$ components will accumulate a dynamic phase. The accumulated phase can be any value in $[0,2\pi]$ by choosing an appropriate detuning $\Delta$:
	
	\begin{equation}
		\begin{split}
			& \ket{\psi_0} = a\ket{10} + b\ket{01} = a'\ket{\Phi_1(0)}+b'\ket{\Phi_2(0)}\,\\
			& \xrightarrow{\mathrm{evolution}} \ket{\psi_t}=a'\ket{\Phi_1(t)}+b'e^{-\ii\Lambda_2(t)}\ket{\Phi_2(t)}\,
		\end{split}
	\end{equation}
	with
	\begin{equation}
		\hspace{-2mm}\Lambda_2(t)=-\int^t_{0}\!\left(\sqrt{\Omega_0(t')^2+\Omega_1(t')^2+(\tfrac{\Delta}{2})^2}-\tfrac{\Delta}{2}\right)\mathrm{d}t'.
	\end{equation}
	
	Choosing $\ket{\Phi_1(0)}$, $\ket{\Phi_2(0)}$ and $\Delta$ allows us to achieve an arbitrary rotation of a location state~\cite{li2021locationqubit}. Conditional location state manipulations is illustrated in Figure~\ref{fig:STIRSAPLocQb}(d). It enables a conditional location state rotation on site 3-4 depending on the location state of site 2. Hence, one can entangle the location states of two adjacent electrons.
	
	The experimental scheme to probe the entanglement polytopes and demonstrate the violation of polytope inequalities~\eqref{eq:polytope-ineq} in the Borland-Dennis case ($N=3$, $d=6$) consist of two steps: the \emph{preparation} of the four states from Table~\ref{tab:states} and the \emph{measurement} of 1-RDM on such states.

	\subsection{Protocol for state preparation}\label{sec:operationsequences}
	In order to prepare the characteristically entangled states, we need to be able to apply certain gate to an initially unentangled state $\ket{\mathrm{Slater}}$. Moreover, we will also need to apply such gates when we read out the natural occupation numbers of the final state.
	
\begin{figure*}[t]
		\centering
		\includegraphics[width=\linewidth]{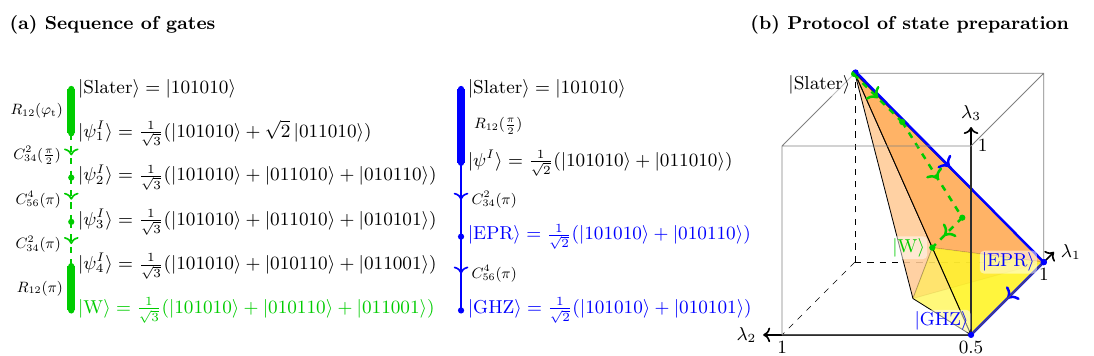}
		\ccaption{State preparation}{\textbf{(a)} We review the details of our protocol to prepare characteristically entangled states. We list all intermediate states and indicate which gate we apply. $R_{ij}(\varphi)$ refers to a \emph{rotation} by $\varphi$ between sites $i$ and $j$, while $C^k_{ij}(\varphi)$ refers to a \emph{controlled rotation} by $\varphi$ between sites $i$ and $j$ that is controlled by site $k$. We indicate by arrows where we move in the polytope, \ie change the natural occupation numbers, while the thick lines denote gates which change the state, but not the natural occupation number, such that they will correspond to a single point in (b) where we show the change of natural occupation numbers. \textbf{(b)} We visualize our protocol in the space of natural occupation numbers.}
		\label{fig:statePrep}
	\end{figure*}

	The Slater state can be prepared as a result of charge initialization~\cite{hastrup2020charging} with electrons occupying sites 1, 3 and 5. The preparation procedures for the entangled fermionic states are illustrated in Figure~\ref{fig:statePrep} and discussed in~\ref{sec:operationsequences}. Preparing such states with high fidelity is indeed an experimental challenge. However numerical studies on all-optical quantum gates~\cite{li2021locationqubit} indicate realistic single qubit gate (two-qubit CNOT gate) fidelities above 0.9999 (0.999). The quantum dot parameters that enter the simulations are presented in section~\ref{sec:simulation}.
	
	The \textbf{rotation gate} $R_{ij}(\varphi)$ acts on the two sites $i$ and $j$, such that
	\begin{linenomath*}\begin{align}
			\begin{split}
				R_{ij}(\varphi)\ket{1_i,0_j}&=\cos{\tfrac{\varphi}{2}}\ket{1_i,0_j}+\sin{\tfrac{\varphi}{2}}\ket{0_i,1_j}\,,\\
				R_{ij}(\varphi)\ket{0_i,1_j}&=\cos{\tfrac{\varphi}{2}}\ket{0_i,1_j}-\sin{\tfrac{\varphi}{2}}\ket{1_i,0_j}\,.
			\end{split}
	\end{align}\end{linenomath*}
	where $\ket{n_i,n_j}$ refers to the occupation numbers on sites $i$ and $j$. Further note that $R_{ij}(\varphi)$ acts as the identity on $\ket{0_i,0_j}$ and $\ket{1_i,1_j}$, as there is no change if both sites or neither site are occupied. For our state protocol, we will be particularly interested in $\varphi=\pi$, which swaps the position of electrons between the sites, and $\varphi=\frac{\pi}{2}$, which generally creates a superposition of electrons on either site. Finally, we will also need the rotation angle $\varphi_{\mathrm{t}}$ satisfying
	\begin{linenomath*}\begin{align}
			\textstyle   \sin(\varphi_{\mathrm{t}})=\sqrt{\tfrac{2}{3}}\quad \Rightarrow\quad \varphi_{\mathrm{t}}=\operatorname{arcsin}(\sqrt{\tfrac{2}{3}})\,.
	\end{align}\end{linenomath*}
	The \textbf{controlled rotation gate} $C_{ij}^k(\varphi)$ is controlled depending on if there is an electron sitting on site $k$. Only if this is the case, $C_{ij}^k(\varphi)$ acts as the rotation gate $R_{ij}(\varphi)$, while otherwise it will not change the state. We therefore have
	\begin{linenomath*}\begin{align}
			C_{ij}^k(\varphi)\ket{n_k,n_i,n_j}=(n_k R_{ij}(\varphi)+(1-n_k)\id)\ket{n_k,n_i,n_j}\,.
	\end{align}\end{linenomath*}
	The rotations $R_{ij}(\varphi)$ and conditional rotations $C_{ij}^k(\varphi)$ are essentially single and two qubit gates. Their experimental implementation is described in greater detail in~\cite{li2021locationqubit}. After initiating the experimental setup with the state $\ket{\mathrm{Slater}}$ consisting of three electrons placed on sites $1$, $3$ and $5$, we can prepare any of the other characteristically entangled state $\ket{\mathrm{W}}$, $\ket{\mathrm{EPR}}$ and $\ket{\mathrm{GHZ}}$ by applying the gates, as elucidated in the state preparation protocol of Figure~\ref{fig:statePrep}.
	
	\subsection{Protocol for measuring natural occupation numbers}\label{sec:measuring-occupation-numbers}
	Conventionally, to gain access to the complete description of a system, a full quantum state tomography can be employed. However, the required number of projective measurements scales exponentially with the number of electrons~\cite{James2001MeasQubits}. For the purpose of probing the entanglement polytopes, it suffices to measure the 1-RDM, as also discussed in~\cite{amiet1999reconstructing,head2019satisfying}. The number of measurements scales quadratically with the number of sites~\cite{walter2013entanglement}, which is in principle much more feasible. The 1-RDM has diagonal and off-diagonal entries. The diagonal entries are the occupation numbers in a given basis that can be measured via state-selective resonant fluorescence (Figure~\ref{fig:STIRSAPLocQb}(c) and section \ref{sec:exp-measure}). The off-diagonal entries can be measured in the same way after a suitable basis transformation, which will be explained in section~\ref{sec:measOffDiag}.

	\subsubsection{Measurement of diagonal entries}\label{sec:exp-measure}
\begin{figure}[t]
		\centering
		\includegraphics[width=\linewidth]{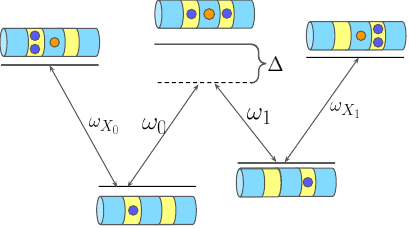}
		\ccaption{The resonant fluorescence energy level diagram}{The left and right most transitions are the one to be exploited in the resonant readout scheme. The charge configurations that are coupled by the transitions are shown in the figure together with the $\Lambda$-system for reference.}
		\label{fig:measurementSetup}
	\end{figure}

	For our crystal-phase quantum dot system we perform state-dependent fluorescence to measure the electronic occupation numbers (diagonal entries of 1-RDM). We describe the concept on a simplified system with only two crystal-phase quantum dots. The setup is a cross-polarization setup for resonant fluorescence as described in~\cite{leandro2020resonant}. The key idea is that if an electron is located at a particular site, we can drive a transition between this charge configuration and a negatively charged exciton state where two electrons are located on the same dot and a hole is located on an adjacent quantum dot for holes as shown in Figure~\ref{fig:measurementSetup}. The transition energies should be designed such that the two outer transitions can be driven resonantly without inducing the two inner transitions that can modify site occupancy.

	By driving the transition resonantly, the quantum dot will begin to emit at the same frequency but in an orthogonal orientation, given that the charge configuration is the one addressed in the transition. The resonant driving energy should be lower than the transition energy between either of the charge states $\ket{10}$ and $\ket{01}$ and their common excited state $\ket{X^-}$. This is to ensure that the state $\ket{X^-}$ is not being populated and induce errors to the measurement. The two transitions $\ket{10}\leftrightarrow \ket{X_0^-}$ and $\ket{01}\leftrightarrow \ket{X_1^-}$ should naturally have different transition energies to ensure state-dependence of the resonant excitation and detection scheme. The fact that the two transitions have different energies requires a resonant excitation frequency for each charge state $\ket{10}$ and $\ket{01}$ respectively, one of the frequencies will be higher than the other, leading to non-resonant excitation of the lower energy transition. But this could be mitigated by spectral filtering of the emitted signal, such that only the emitted signal at the same frequency as the excitation laser will be registered.
	
	The energy of the charge configurations and hence the transition energies can be engineered to some extent by manipulating the sample parameters such as material composition (GaAs or InP) nanowire diameter, geometry (hexagonal or cylindrical wire) and quantum dot lengths. On each measurement trial, we will either get an emitted phonon at the resonant excitation energy or no signal. By repeating the trials at a sufficiently high rate, the count rate for resonant photons collected will be high for the charge state of interest.

	\subsubsection{Measurement of off-diagonal entries}
	\label{sec:measOffDiag}
	In order to compute the natural occupation numbers, we need to measure the matrix entries $\gamma_{ij}$ of the 1-RDM. For a system of six sites, this is a $6$-by-$6$ Hermitian matrix subject to the constraint $\mathrm{Tr}(\gamma)=N$, so we will generally have $35$ independent entries. Our measurement setup is able to extract diagonal entries (electronic on-site occupation numbers $n_i$), so we will need to apply additional gates to find off-diagonal entries.
	
	We focus on two fixed site indices $i$ and $j$, such that
	\begin{linenomath*}\begin{align}
			\begin{pmatrix}
				\gamma_{ii} & \gamma_{ij}\\
				\gamma_{ij}^* & \gamma_{jj}
			\end{pmatrix}=\begin{pmatrix}
				\alpha & x+\ii y\\
				x-\ii y & \beta
			\end{pmatrix}\,.
	\end{align}\end{linenomath*}
	In order to measure the off-diagonal elements of the 1-RDM, $\gamma_{ij} = x+\ii y$, which potentially has a real and imaginary part, we need to act with an additional transformation. The purpose of the transformations are to transfer the information of the off-diagonals onto the diagonals which is directly measurable. The simplest setup are the following two types of Bogoliubov transformations $\hat{a}_i\to\hat{b}_i$ or $\hat{a}_i\to\hat{c}_i$, respectively:
	\begin{linenomath*}\begin{equation}
			\begin{aligned}
				\hat{b}_i&=\tfrac{1}{\sqrt{2}}(\hat{a}_i+\hat{a}_j)\,,&\hat{c}_i&=\tfrac{1}{\sqrt{2}}(\hat{a}_i+\ii\hat{a}_j)\,,\\
				\hat{b}_j&=\tfrac{1}{\sqrt{2}}(\hat{a}_i-\hat{a}_j)\,,&\hat{c}_j&=\tfrac{1}{\sqrt{2}}(\hat{a}_i-\ii\hat{a}_j)\,.
			\end{aligned}
	\end{equation}\end{linenomath*}
	In order to perform the respective operations, we need to apply either the rotation $R_{ij}(\tfrac{\pi}{2})$ for $\hat{b}$ or a the $R_{ij}(\tfrac{\pi}{2})$ plus another $\frac{\pi}{2}$ rotation with respect to the axis $(0,0,1)$ as explained in~\cite{caillet2007precision} for $\hat{c}$. If $i$ and $j$ are not adjacent, we can first bring the respective sites together by applying a sequence of swaps $R_{i,i+1}(\pi)$ until we reach site $j-1$, where we assumed $i<j$ without loss of generality. The two transformations lead to the transformed 1-RDM $\gamma^b_{ij}=\braket{\hat{b}_i^\dagger\hat{b}_j}$ and $\gamma^c_{ij}=\braket{\hat{c}_i^\dagger\hat{c}_j}$, which are given by
	\begin{linenomath*}\begin{align}
			\begin{split}
				\gamma^{b}&=\frac{1}{2}\begin{pmatrix}
					\alpha+\beta+2x & \alpha-\beta-2\ii y\\
					\alpha-\beta+2\ii y & \alpha+\beta-2x
				\end{pmatrix}\,,\\
				\gamma^c&=\frac{1}{2}\begin{pmatrix}
					\alpha+\beta-2y & \alpha-\beta-2\ii x\\
					\alpha-\beta+2\ii x & \alpha+\beta+2y
				\end{pmatrix}\,.
			\end{split}\label{eq:gamma-bc}
	\end{align}\end{linenomath*}
	We can then extract both $x$ and $y$ from the measured diagonal elements.
	
	As a side note, an off-diagonal entry might couple sites that are not neighbouring sites (\eg site 1 and 6), so a series of $\pi$-rotations are needed to first bring the sites of interest together (to site 3-4).
	
	\section{Numerical simulation}\label{sec:simulation}
\begin{figure*}[t]
		\includegraphics[width=\linewidth]{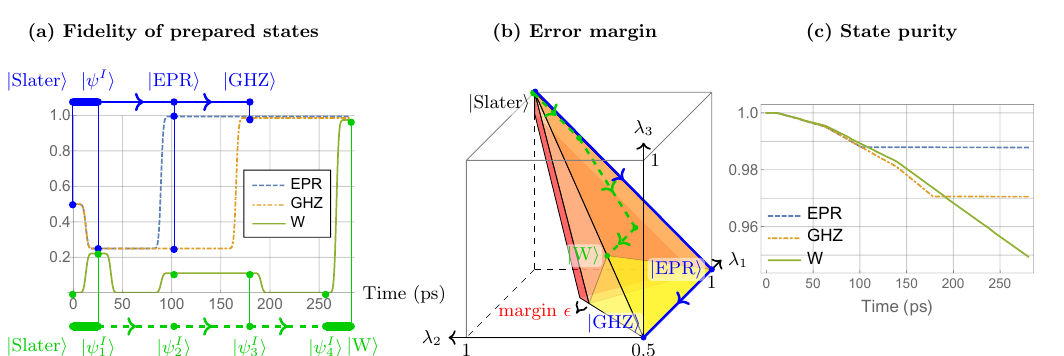}
		\ccaption{Numerical simulations}{\textbf{(a)} The fidelities of the states $\ket{\mathrm{EPR}}$ (dashed), $\ket{\mathrm{GHZ}}$ (dot-dashed), $\ket{\mathrm{W}}$ (full) during their respective preparation procedure and subsequent free evolution. The $\ket{\mathrm{Slater}}$ state is assumed to be initialized with 100\% fidelity. The fidelities of the prepared states are calculated relative to their respective pure states. The states $\ket{\psi^I}$, $\ket{\psi^I_1}$ ... $\ket{\psi^I_4}$ are intermediate states during state preparation well described in~\ref{sec:operationsequences}. The blue~(top) and green~(bottom) traces correspond to traces of the same color in \textbf{(b)}. \textbf{(b)} We show the margin (red) of the extended Pauli principle if the global state is slightly mixed. More precisely, we consider a global state $\rho$ with the spectrum $(1-\epsilon,\epsilon_1,\dots,\epsilon_{n-1})$ with small $\epsilon=\sum_{i}\epsilon_i$, for which the inequality of the extended Pauli principle~\eqref{eq:schematic} becomes $\lambda_1+\lambda_2-\lambda_3\leq 1+\epsilon$, as shown in~\eqref{eq:BD-mixed}. The error margin $\epsilon$ is set to 0.06. \textbf{(c)} The purity of the prepared states $\ket{\mathrm{EPR}}$ (dashed), $\ket{\mathrm{GHZ}}$ (dot-dashed), $\ket{\mathrm{W}}$ (full) throughout their respective preparation procedure.}
		\label{fig:fidelityTrajectory}
	\end{figure*}

	The state preparation fidelity and purity are simulated (using appendix in~\cite{li2021locationqubit}) and shown in Figure~\ref{fig:fidelityTrajectory}(a,c). We can achieve a fidelity of above 0.97 and a purity of above 0.95 for all states. In the presence of experimental imperfections and limitations of the manipulation scheme~\cite{caillet2007precision}, additional fermionic spin-orbitals will be involved, effectively expanding the system to a higher number of sites. By considering the mixed nature of the states, we derive in section~\ref{sec:slightlymixedpolytope} a weakened Borland-Dennis relation.
	\begin{linenomath*}\begin{align}
			\lambda_1 + \lambda_2 - \lambda_3 \leq 1 + \epsilon
	\end{align}\end{linenomath*}
	The simulated states are all subject to the weakened relation (Figure~\ref{fig:fidelityTrajectory}(b)). In Table~\ref{tab:merits} we present the figures of merit simulated for each of the states together with the error estimates (section \ref{sec:errors}). In section~\ref{sec:purity}, a procedure to prepare the state then rewind the preparation expecting the system to return to the initial state (Loschmidt echo~\cite{shaffer2021loschmidt}) is simulated to quantify the purity of the prepared states.
	
	\begin{table*}[t!]
		\centering
		\renewcommand{\arraystretch}{1.3}
		\begin{tabular}{lcccccc}
			\textbf{State} & \textbf{Natural occupation numbers $\vec{\lambda}$} & \textbf{Fidelity} & \textbf{Error margin $\epsilon$} & \textbf{Merit function $F_1$}  & \textbf{Merit function $F_2$}   \\ \hline
			$\ket{\mathrm{EPR}}$ & (1.0000, 0.5012, 0.5011,0.4988, 0.4987, 0) & 0.9939 & 0.0061 & 1.0001 & 2       \\
			$\ket{\mathrm{GHZ}}$ & (0.5013, 0.5011, 0.5003, 0.4995, 0.4987, 0.4987) & 0.9850 & 0.0149 & 0.5013 & 1.5019  \\
			$\ket{\mathrm{W}}$   & (0.6670, 0.6666, 0.6659, 0.3335, 0.3332, 0.3330) & 0.9741 & 0.0258 & 0.6676 & 1.66705
		\end{tabular}
		\ccaption{Figures of merit}{The columns are the natural occupation numbers, state preparation fidelities, the error margin and the two Borland-Dennis parameters $F_1 = \lambda_1 + \lambda_2 - \lambda_3$ and $F_2 =  \lambda_1 + \lambda_2 + \lambda_4$ that are expected to satisfy  $F_1 \leq 1 + \epsilon$ and $F_2 \leq 2 + \epsilon$. The state preparation fidelities are the final fidelities as shown in figure~\ref{fig:fidelityTrajectory}\textbf{(a)} around $t=280$ps. We find good agreement with the simulation.
		}
		\label{tab:merits}
	\end{table*}
	
	As discussed, the proposed experimental platform is a system of multiple crystal-phase quantum dots. More precisely, in the numerical simulations, InP nanowires with a diameter of 50nm, length of 1$\mu$m are used. The main composition of a nanowire is InP in the wurtzite phase while crystal-phase quantum dots are defined as short zincblende sections ($20$nm), individual quantum dots are separated by longer wurtzite sections (50-100nm). The quantum dot parameters such as InP material parameters, dimensions of the individual dots and separation between the dots relevant for the evaluation of spontaneous emission rate and dephasing rate. The spontaneous emission rate and dephasing rate are later used in the calculation of single-qubit and two-qubit gate fidelities. Dephasing rate due to electron-phonon deformation coupling are calculated based on with a single crystal phase quantum dot with 20nm length over a range of temperatures. At 4K, the dephasing rate is $1.66\cdot10^5$ s$^{-1}$. Spontaneous emission rates are calculated in a structure with a pair of crystal phase quantum dots (ZB insertions) that defines a location qubit. The first quantum dot varies from $20$ to $50$nm in length, while the right quantum dot is fixed at $19$nm. The WZ section that defines the separation between the quantum dots are varied between 50 and 200nm. The minimal (maximal) spontaneous emission rate is found to be $9.27027\cdot10^7$ s$^{-1}$ ($7.79388\cdot10^8$ s$^{-1}$). The minimum is obtained at ZB $42$nm, WZ $200$nm, while the maximum is obtained at ZB $36$nm, WZ $70$nm. Any other dimensions within the range, including the dimensions that produce the dephasing rate, will yield a spontaneous emission rate in between. In the numerical simulations the maximal spontaneous emission rate is used to produce an underestimate of the gate fidelities and state preparation fidelities. Details on the methods are available in~\cite{li2021locationqubit}.
	
	Our multi-quantum-dot system is in essence a nanostructure where electrons can be selectively and deterministically loaded~\cite{hastrup2020charging}. This allows the preparation of any charge state of the type $\ket{n_1 n_2 \dots n_N}$ where $n_i\in\{0,1\}$ and $N$ is the total number of quantum dots in a nanowire. The number of electrons and on which quantum dots they localize is not restricted. The system is qubitized in the current proposal since a scheme to coherently manipulate charge states is readily available~\cite{li2021locationqubit}. However, this does not prevent the preparation of any other charge states. Our system is intrinsically fermionic suitable for the testing of the extended Pauli constraints. Experimentally we can access and manipulate other charge states such as $\ket{110010}$ in the exact same way as charge states $\ket{10}$ and $\ket{01}$ within a pair of neighbouring quantum dots.
	
	\begin{figure*}[t]
		\centering
		\includegraphics[width=\linewidth]{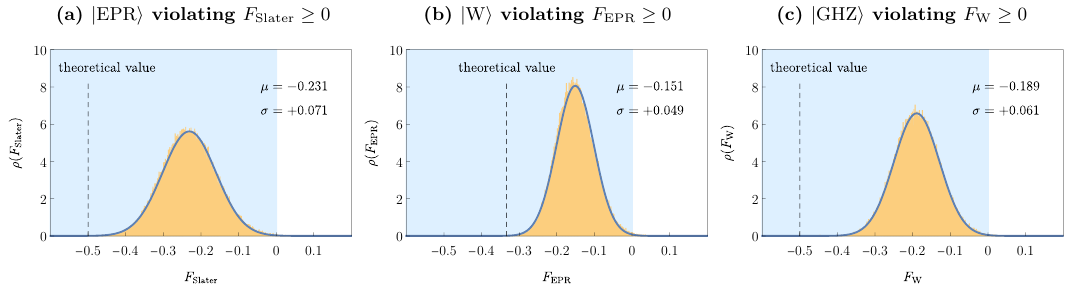}
		\vspace{-4mm}
		\ccaption{Error estimates}{
			We determine the largest possible standard deviation $\epsilon$ for the matrix entries of the 1-RDM, such that we can still confirm a violation (bright blue) of the respective inequalities $F\geq 0$, while $F$ is allows to largely differ from its theoretical value~\eqref{eq:th-values} (dashed line). We find the errors listed in~\eqref{eq:error} for a $99.9\%$ certainty of violation. We also indicate expectation value $\mu$ and standard deviation $\sigma$ of the $F$ (which is of course different from the standard deviation $\epsilon$ of the matrix entries).
		}
		\label{fig:errors}
	\end{figure*}
	
	\section{Error estimates}\label{sec:errors}
	Our probe of the extended Pauli principle crucially relies on being able to measure the natural occupation numbers of arbitrary states and in particular of the characteristically entangled states to sufficient accuracy. In particular, we need to be able to show that the respective polytope constraints~\eqref{eq:polytope-ineq} are violated.
	
	We therefore first determine what the maximally allowed error for the matrix entries of the 1-RDM is, so that we can still confirm a violation with $3\sigma$, \ie such that we only expect $0.1\%$ of random samples to \emph{not} violate the inequalities. For this, we can define the following three merit functions based on the polytope inequalities~\eqref{eq:polytope-ineq} which we will violate with respective next states:
	\begin{linenomath*}\begin{align}
			\begin{split}
				F_{\mathrm{Slater}}&=\lambda_2-1\geq 0\,,\\
				F_{\mathrm{EPR}}&=\lambda_1-1\geq 0\,,\\
				F_{\mathrm{W}}&=2-\lambda_1-\lambda_2-\lambda_3\geq 0\,,
			\end{split}\label{eq:polytope-merits}
	\end{align}\end{linenomath*}
	were we defined each merit function in such a way that the respective inequality requires the entries to be equal or larger than zero. We then simulate the histograms $\rho(F)$ for $10^5$ random samples, where we assume that the matrix entries of the 1-RDM are distributed as Gaussians with standard deviation $\epsilon$ around their theoretical values, \ie their expected values for ideal pure states
	\begin{linenomath*}\begin{align}
			F_{\mathrm{Slater}}(\ket{\mathrm{EPR}})=F_{\mathrm{W}}(\ket{\mathrm{GHZ}})=-\tfrac{1}{2}\,,\, F_{\mathrm{EPR}}(\ket{\mathrm{W}})=-\tfrac{1}{3}\,.\label{eq:th-values}
	\end{align}\end{linenomath*}
	We assume that the 1-RDM is Hermitian with independently distributed real and imaginary entries. Based on this, we find the following maximal errors (standard deviations) of its matrix values\footnote{In the special case of $\ket{\mathrm{EPR}}$, the 1-RDM has the entry $\gamma_{66}=0$, so Gaussian distributed perturbations may lead to an unphysical negative value. We therefore took the absolute value of the Gaussian distributed perturbation to ensure that all diagonal entries are positive.} (see figure~\ref{fig:errors})
	\begin{linenomath*}\begin{align}
			\epsilon_{\mathrm{EPR}}=0.083\,,\quad\epsilon_{\mathrm{W}}=0.055\,,\quad\epsilon_{\mathrm{GHZ}}=0.037\label{eq:error}
	\end{align}\end{linenomath*}
	for the respective states. Our code is public via~\cite{github}.
	
	Let us now analyze if we can stay below these maximal errors. While it is difficult to take all sources of error into account, we model the two types of error, which we expect to dominate in the experimental realization. These two errors are (a) decoherence due to spontaneous emission and dephasing as reviewed in~\cite{li2021locationqubit} and (b) the statistical error of measuring expectation values from a finite sample.
	
	\textbf{(a) Decoherence.} The matrix entries of the 1-RDM are measured based on finding the expectation values $\braket{\hat{n}_i}$ for the respective state $\rho$ (diagonal entries) or for a transformed state $\rho'$ (off-diagonal entries), where we need to apply additional gates as explained in section~\ref{sec:measuring-occupation-numbers}. The simulations reviewed in~\cite{li2021locationqubit} give 1-RDM errors of the order of $10^{-3}$ and smaller. For this, we calculated 1-RDM for the respective final states and compared them with the theoretical prediction.
	
	\textbf{(b) Statistical error.} As explained in section~\ref{sec:exp-measure}, all measurements of the 1-RDM are reduced to measuring the expected occupation $\braket{\hat{n}}$ at a site $i$, so each measurement will either yield $1$ (occupied) or $0$ (unoccupied). The actual measurement outcomes will be distributed according to a binomial distribution with standard deviation $\sigma=\sqrt{\braket{\hat{n}}(1-\braket{\hat{n}})/M}$, where $M$ is the number of measurements performed, which is bounded from above by $1/\sqrt{4M}$. By performing $M\approx 10^5$ measurements, we expect an error of ca. $10^{-3}$.
	
	In summary, we find simulated errors that are at least an order of magnitude smaller than the maximally allowed values listed in~\eqref{eq:error} for a confidence of $99.9\%$, so we expect that our proposed experimental setup allows us to confirm violation of polytope inequality to high accuracy.
	
	\section{Pure vs. mixed states}\label{sec:pure-vs-mixed}
	The Borland-Dennis relations are derived for pure fermionic states. Since experimental noise and decoherence is inevitable, the states being prepared in an experiment will always be mixed to some extend. The experimental demonstration of Borland-Dennis relations therefore relies on high purity of the entangled states. Therefore, it is an important part of the experiment to measure the purity of the total quantum state in the system, as explained in section~\ref{sec:purity}. We further determine in section~\ref{sec:slightlymixedpolytope} how the Borland-Dennis inequalities change for a \emph{slightly} mixed states leading to a small additional error margin, as already indicated in figure~\ref{fig:fidelityTrajectory}.
	
	\subsection{Quantifying purity}\label{sec:purity}
	The purity of a quantum state with density operator $\rho$ is defined as $\operatorname{tr}\rho^2$. The purity of an experimentally prepared multi-particle entangled state can be estimated with the ``entangle-disentangle'' procedure as described in~\cite{walter2013entanglement}, which is formally known as Loschmidt echo~\cite{shaffer2021loschmidt} in the quantum simulation community. The key idea is that if the system is initially prepared in an eigenstate, by implementing a forward time evolution followed by a backward time evolution that is the conjugate of the former, the system is expected to return to the initial state. This process is ``a way of increasing the confidence in the (quantum simulation) result"~\cite{cirac2012a}, with one of the first experimental implementations being the Hahn echo experiment~\cite{hahn1950spin}. First suppose that the ideal state preparation unitary operation $U$ (entangling operation) is approximated by a lossy quantum channel $C_1$ and the resultant entangled state is described by the density operator $\rho=C_1(\rho_{init})$, where $\rho_{init}$ is the density operator for an initial Slater state. Another lossy quantum channel $C_2$ approximates the inverse of the entangle operation $U^{-1}$ (disentangling operation) with the resultant density operator $\rho'=C_2(\rho)$. If both quantum channels are not deviating too far from the ideal operations, $\rho'$ will be approximately $\rho_{init}$. At the same time, if the purity of a quantum state does not increase in a lossy quantum channel, the purity of $\rho'$ will give a lower bound of the purity of $\rho$, which is the state of interest. It is shown~\cite{walter2013entanglement} that a lower bound $p$ of the purity of $\rho$ is given as
	\begin{linenomath*}\begin{align}
			\mathrm{tr}\rho^2 \geq \mathrm{tr}\rho'^2 \geq \sum^N_{k=1} \norm{ \vec{\lambda}^{k}(\rho') }_2^2 - (N-1) = p. \label{eq:purity}
	\end{align}\end{linenomath*}
	In the context of Borland-Dennis $N=3$. $\vec{\lambda}^{k}(\rho')$ are vectors consisting of the natural occupation numbers of the $k$'th pair of quantum dots. Each single particle density operator $\rho'^{(k)}$ can be tomographically measured under local operations. The number of measurements would therefore scale linearly with the number of fermions. 
	
\begin{figure}[t]
		\centering
		\includegraphics[width=\linewidth]{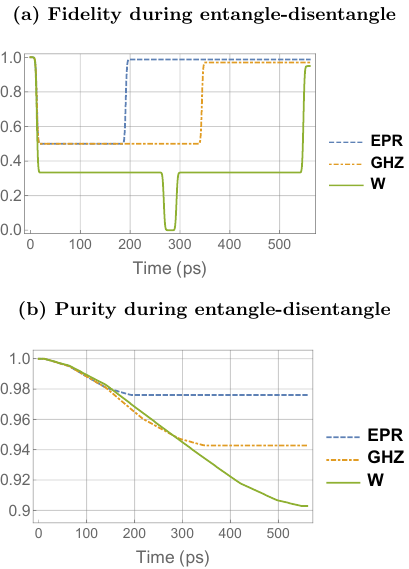}
		\ccaption{Fidelity and purity of the states during the entangle-disentangle procedure}{The time evolution of the \textbf{(a)} fidelity and \textbf{(b)} purity of the prepared states $\ket{\mathrm{EPR}}$ (dashed), $\ket{\mathrm{GHZ}}$ (dot-dashed), $\ket{\mathrm{W}}$ (full). The fidelities are calculated with respect to the Slater state, since after the procedure the states are expected to return to the Slater state. As a result, the fidelities start from 100\% for all states and traverse through the entangle-unentangle procedure. }
		\label{fig:purity}
	\end{figure}

	However, not all decoherence processes decrease the state purity. For instance, the purity of an initial even superposition $\ket{\psi_i} = \frac{1}{\sqrt{2}} (\ket{10}+\ket{01})$ is decreased by the process of pure dephasing and gradually evolves towards a statistical mixture $\rho = \frac{1}{2} (\ket{10}\bra{10} + \ket{01}\bra{01})$. But the same initial state would regain a purity of 1 in the long term limit if the decoherence process is relaxation such that the population in the state $\ket{01}$ is gradually transferred to $\ket{10}$. It is therefore essential for the entangle-disentangle procedure that the dominant decoherence processes decrease the purity of a state within the time frame of interest. As explained in~\cite{li2021locationqubit}, for the charge states on crystal-phase quantum dots, electron-phonon interaction and more specifically deformation potential coupling with the longitudinal acoustic phonons is the dominant physical mechanism for decoherence~\cite{li2021locationqubit}. Relaxation (tunneling) can be effectively eliminated by separating the dots sufficiently far apart (typically $>$ 80nm). At a typical cryogenic temperature $T$=4K, a separation of 100nm between two dots of 20nm and 19nm respectively corresponds to a tunneling lifetime of 62ms. For the same dots at the same temperature, dephasing time $T_2$ is 6$\mu$s, which is much more considerable~\cite{li2021locationqubit}. Another source of decoherence is spontaneous emission from the common excited state $\ket{X^-}$ to the two charge states $\ket{10}$ and $\ket{01}$. But since the common excited state is populated only during the charge state operations and the component is meant to be tiny, the effect of spontaneous emission does not affect the long term coherence of the state. It is therefore valid to use the ``entangle-disentangle'' procedure to examine the purity of the experimentally prepared state in crystal-phase quantum dots. The simulated fidelity and purity as a function of time are given in Figure~\ref{fig:purity}. It can be seen that the impurity found with this method is roughly doubled. Yet the entangle-disentangle procedure for purity estimation is experimentally feasible since it only requires one projective measurement to the Slater state and no further complication beyond state preparation procedures. The final purities are simulated to be 0.9761, 0.9428 and 0.9028 for each of the states $\ket{\mathrm{EPR}}$, $\ket{\mathrm{GHZ}}$ and $\ket{\mathrm{W}}$, serving well as the lower bound estimate for the purities.
	
	\subsection{Extended Pauli polytope for slightly mixed states}
	\label{sec:slightlymixedpolytope}
	
	The constraints of the extended Pauli principle only apply to pure states. There are no further constraints on the natural occupation numbers $\lambda_i$ except the regular Pauli principle, \ie $0\leq\lambda_i\leq 1$ (and our choice of sorting $\lambda_1\geq\dots\geq\lambda_d$), if we allow for arbitrarily mixed states. However, if we can put constraints on \emph{how mixed} a given state is, one can derive error bars around the extended Pauli principle.
	
	\begin{proposition}
		The natural occupation numbers $1\geq\lambda_1\geq\dots\geq\lambda_6\geq 0$ of a mixed state $\rho$ in the Borland-Dennis system $\Lambda^3\mathcal{H}_6$ whose largest eigenvalue is given by $(1-\epsilon)$ satisfies the slightly weakened Borland-Dennis conditions
		\begin{linenomath*}\begin{align}
				\begin{split}
					\lambda_1+\lambda_2-\lambda_3&\leq 1+\epsilon\,,\\
					\lambda_1+\lambda_2+\lambda_4&\leq 2+\epsilon\,.
				\end{split}\label{eq:BD-mixed}
		\end{align}\end{linenomath*}
		Note that these equations are now independent, as we do not have the constraints $\lambda_1+\lambda_6=\lambda_2+\lambda_5=\lambda_3+\lambda_4=1$ anymore (only $\sum^6_{i=1}\lambda_i=3$).
	\end{proposition}
	\begin{proof}
		Based on the assumptions, we can write
		\begin{linenomath*}\begin{align}
				\rho=(1-\epsilon)\ket{\psi_0}\bra{\psi_0}+\epsilon \rho_1\,,
		\end{align}\end{linenomath*}
		where $\ket{\psi_0}$ is the normalized eigenvector of $\rho$ with eigenvalue $(1-\epsilon)$ and $\rho_1$ is an arbitrary mixed state, such that $\rho_1\ket{\psi_0}=0$. We have the 1-RDM $\gamma=(1-\epsilon) \gamma_0+\epsilon\gamma_1$ with
		\begin{linenomath*}\begin{align}
				\gamma_0^\psi=\braket{\psi_0|\hat{a}^\dagger_j\hat{a}_i|\psi_0}\quad\text{and}\quad\gamma_1=\mathrm{Tr}(\rho_1\hat{a}^\dagger_j\hat{a}_i)\,.
		\end{align}\end{linenomath*}
		Let us denote the ordered eigenvalues of $\gamma_1$ by $\lambda_i^\psi$, which satisfy the Borland-Dennis conditions
		\begin{linenomath*}\begin{align}
				\begin{split}
					\lambda^\psi_1+\lambda^\psi_2+\lambda^\psi_4&\leq 2\,,\\
					\lambda^\psi_1+\lambda^\psi_6=\lambda^\psi_2+\lambda^\psi_5=\lambda^\psi_3+\lambda^\psi_4&=1\,.\label{eq:BD-in}
				\end{split}
		\end{align}\end{linenomath*}
		For the eigenvalues of $\gamma_1$, we only know that they satisfy the regular Pauli principle, \ie lie in the interval $[0,1]$, so that adding $\epsilon\gamma_1$ can only increase the eigenvalues, but at most by $\epsilon$. From these considerations, it follows immediately that the ordered eigenvalues $\lambda_i$ of $\gamma$ satisfy
		\begin{linenomath*}\begin{align}
				(1-\epsilon)\lambda_i^\psi\leq \lambda_i\leq(1-\epsilon)\lambda_i^\psi+\epsilon\,.
		\end{align}\end{linenomath*}
		Using the fact that $\lambda_i^\psi$ satisfies~\eqref{eq:BD-in} then implies
		\begin{linenomath*}\begin{align}
				\begin{split}
					\hspace{-3mm}\lambda_1\!+\!\lambda_2\!+\!\lambda_4&\!\leq\! (1\!-\!\epsilon)(\underbrace{\lambda_1^\psi\!+\!\lambda_2^\psi\!+\!\lambda_4^\psi}_{=2})+3\epsilon=2\!+\!\epsilon\,,\\
					\hspace{-3mm}\lambda_1\!+\!\lambda_2\!-\!\lambda_3&\!\leq\! (1\!-\!\epsilon)(\underbrace{\lambda_1^\psi\!+\!\lambda_2^\psi\!-\!\lambda_3^\psi}_{=1})+2\epsilon=1\!+\!\epsilon\,,
				\end{split}
		\end{align}\end{linenomath*}
		where used $\lambda_i\leq \lambda_i^\psi+\epsilon$ and $\lambda_3\geq \lambda_3^\psi$.
	\end{proof}
	
	We also tested the inequalities~\eqref{eq:BD-mixed} numerically. We implemented a simple random optimization algorithm that initiates on some random mixed state
	\begin{linenomath*}\begin{align}\label{eq:rho-form}
			\rho=(1-\epsilon)\ket{\psi_0}\bra{\psi_0}+\epsilon \rho_1\,.
	\end{align}\end{linenomath*}
	We then randomly perturb the states $\ket{\psi_0}$ and $\rho_1$ to find a new mixed state $\rho'$ for which $f_1(\rho')=\lambda_1+\lambda_2-\lambda_3$ or $f_2(\rho')=\lambda_1+\lambda_2+\lambda_4$ increases. By repeating this process we can stochastically scan the space of density operators of the form~\eqref{eq:rho-form}. This allowed us to find states that are arbitrary close to saturating either of the two inequalities~\eqref{eq:BD-mixed}. In fact, we can even give explicitly the state $\rho=(1-\epsilon)\ket{\psi_0}\bra{\psi_0}+\epsilon\ket{\psi_1}\bra{\psi_1}$ with
	\begin{linenomath*}\begin{align}
			\ket{\psi_0}=\ket{111000}\quad\text{and}\quad\ket{\psi_1}=\ket{110100}\,,
	\end{align}\end{linenomath*}
	whose natural occupation numbers $\vec{\lambda}=(1,1,1-\epsilon,\epsilon,0,0)$ saturate both inequalities.
	
	\section{Discussion}
	Understanding the role of the constrains imposed by the extended Pauli principle has fundamental importance on simulation of fermionic systems, such as the electronic structure of molecules~\cite{reiher,o2016scalable,reiher2017elucidating,mcardle2020quantum}. Since the correlations of electrons are quantum-mechanical in nature, this poses a challenge for our current computing infrastructure, which only operates with classical states. Our proposed probe of the extended Pauli principle and thus of fermionicity itself can therefore also be regarded as a step towards quantum simulations of molecules. Our approach is also general, as the proposed probes can be used in different, as well as larger, physical platforms, including trapped fermionic atoms~\cite{murmann2015two}. To date, the extended Pauli principle remains to be extensively explored, but only theoretically, with no way to be verified experimentally. Our array of crystal-phase quantum dots in a nanowire is thus the pathway for experimental exploration.
	
	\begin{acknowledgments}
		MC thanks Selim Jochim for early discussions on the topic. We thank Robin Reuvers for comments on the draft. MC acknowledges financial support from the VILLUM FONDEN via the QMATH Centre of Excellence (Grant No.10059), the European Research Council (ERC Grant Agreement No. 81876) and the Novo Nordisk Foundation (grant NNF20OC0059939 ‘Quantum for Life’). LH gratefully acknowledges support by the Alexander von Humboldt Foundation. NA acknowledges financial support from the European Research Council (ERC Grant Agreement No. 101003378).
	\end{acknowledgments}

	\bibliography{references}
	
\end{document}